\documentclass[epsf,12pt,russian]{article}

\usepackage{amssymb}

\usepackage[dvips]{graphicx}

\begin{document}

\title
{Summation of diagrams in $N=1$ supersymmetric electrodynamics,
regularized by higher derivatives.}

\author{K.V.Stepanyantz}

\maketitle

\begin{center}

{\em Moscow State University, physical faculty,\\
department of theoretical physics.\\
$119992$, Moscow, Russia}

\end{center}

\begin{abstract}
For the massless $N=1$ supersymmetric electrodynamics, regularized
by higher derivatives, the Feynman diagrams, which define the
divergent part of the two-point Green function and can not be
found from Schwinger-Dyson equations and Ward identities, are
partially summed. The result can be written as a special identity
for Green functions.
\end{abstract}


\section{Introduction}
\hspace{\parindent}

Current indirect proofs of existence of the supersymmetry in the
Standard model make the problem of calculation of quantum
corrections in supersymmetric theories especially urgent. The
supersymmetry essentially improves the ultraviolet behavior of a
theory. Thanks to that it is possible to suggest the form of the
$\beta$-function exactly to all orders of the perturbation theory
even in theories with unextended supersymmetry. The form of the
exact $\beta$-function was proposed first in Ref.
\cite{NSVZ_Instanton} from the investigation of the instanton
contributions structure. For $N=1$ supersymmetric electrodynamics,
which will be considered in this paper, this $\beta$-function
(that is called the exact Novikov, Shifman, Vainshtein and
Zakharov (NSVZ) $\beta$-function) is

\begin{equation}\label{NSVZ_Beta}
\beta(\alpha) = \frac{\alpha^2}{\pi}\Big(1-\gamma(\alpha)\Big),
\end{equation}

\noindent where $\gamma(\alpha)$ is the anomalous dimension of the
matter superfield.

Explicit calculations, made with the dimensional reduction
\cite{Siegel}, confirm this proposal, but require a special choice
of the subtraction scheme \cite{Tarasov,North}. Explicit
calculations in two- \cite{hep,tmf2}, three- \cite{ThreeLoop} and
partially four-loop \cite{Pimenov} approximations  for the $N=1$
supersymmetric electrodynamics with the higher derivative
regularization \cite{Slavnov,Bakeyev} reveal that renormalization
of the operator $W_a C^{ab} W_b$ is exhausted at the one-loop and
the Gell-Mann-Low function coincides with the exact NSVZ
$\beta$-function and has corrections in all orders of the
perturbation theory.

(In the supersymmetric electrodynamics the Gell-Mann-Low
$\beta$-function is defined as follows: If terms, quadratic in the
superfield $V$, in the renormalized effective action in the zero
mass limit has the form

\begin{equation}
\Gamma^{(2)}_V = - \frac{1}{16\pi} \int
\frac{d^4p}{(2\pi)^4}\,V(-p)\,\partial^2\Pi_{1/2} V(p)\,
d^{-1}(\alpha,\mu/p),
\end{equation}

\noindent where $\partial^2\Pi_{1/2}$ is a supersymmetric
transversal projector, the Gell-Mann-Low function is defined by

\begin{equation}\label{GL_Function}
\beta\Big(d(\alpha,x)\Big) \equiv -\frac{\partial}{\partial\ln
x}\, d(\alpha,x)\Bigg|_{x=1}.)
\end{equation}

Nevertheless, the investigation of the supersymmetric
electrodynamics was so far restricted by the three-loop
approximation. It would be interesting to find out if it is
possible to make explicit calculations exactly to all orders of
the perturbation theory. As it was shown in Ref. \cite{SD}, a
large number of Feynman diagrams can be calculated exactly to all
orders of the perturbation theory using the Schwinger-Dyson
equations and the Ward identities. Nevertheless, some diagrams
were not found by this way. A special proposal about the structure
of the remaining contributions was made in Ref. \cite{SD}. It
assumes the existence of an identity, which is not reduced to Ward
identities and can be graphically written as equality to 0 of the
diagram, presented in Fig. \ref{Figure_Undefined_Diagram}, where
the notation $(I_1)^a$ is explained below in the paper. (In the
analytic form this identity is given by Eq. (\ref{New_Identity})).
This identity is nontrivial for non-planar diagrams starting from
the three-loop approximation. In this paper we try to find a way
to prove it exactly to all orders of the perturbation theory.

We consider the massless theory and a sufficiently large class of
Feynman diagrams, which have the only loop of the matter
superfields and remain connected after any two cuts of this loop.
(Such a choice is made because in this case the technique of
calculations is simpler.) For these diagrams the identity, pointed
above, is completely proved in this paper. The proposed method of
the proof seems can be used in the general case, but with some
more technical difficulties.

The paper is organized as follows:

In Sec. \ref{Section_SUSY_QED} the basic information about $N=1$
supersymmetric electrodynamics and its regularization by higher
derivatives is reminded. The Schwinger-Dyson equations for the
considered theory, constructed in Ref. \cite{SD}, are presented in
Sec. \ref{Section_SD}. Calculation of contributions to the
two-point Green function of the gauge field, which can not be
defined from the Ward identities, is made in Sec.
\ref{Section_Green}. (Such contributions of diagrams of the
considered class are proved here to be 0 in the massless case.) In
Sec. \ref{Section_Anomaly_Puzzle} the two-point Green function of
the gauge field is constructed on the basis of the obtained
results. Divergences in it are shown to exist only in the one-loop
approximation, while the Gell-Mann-Low function coincides with the
exact NSVZ $\beta$-function. The obtained results are discussed in
the Conclusion.


\section{$N=1$ supersymmetric electrodynamics and its
regularization by higher derivatives} \label{Section_SUSY_QED}
\hspace{\parindent}

The massless $N=1$ supersymmetric electrodynamics with the higher
derivatives term in the superspace is described by the following
action:

\begin{equation}\label{Regularized_SQED_Action}
S = \frac{1}{4 e^2} \mbox{Re}\int d^4x\,d^2\theta\,W_a C^{ab}
\Big(1+ \frac{\partial^{2n}}{\Lambda^{2n}}\Big) W_b +
\frac{1}{4}\int d^4x\, d^4\theta\, \Big(\phi^* e^{2V}\phi
+\tilde\phi^* e^{-2V}\tilde\phi\Big).
\end{equation}

\noindent Here $\phi$ and $\tilde\phi$ are the chiral matter
superfields and $V$ is a real scalar superfield, which contains
the gauge field $A_\mu$ as a component. The superfield $W_a$ is a
supersymmetric analog of the stress tensor of the gauge field. In
the Abelian case it is defined by

\begin{equation}
W_a = \frac{1}{16} \bar D (1-\gamma_5) D\Big[(1+\gamma_5)D_a
V\Big],
\end{equation}

\noindent where

\begin{equation}
D = \frac{\partial}{\partial\bar\theta} -
i\gamma^\mu\theta\,\partial_\mu
\end{equation}

\noindent is a supersymmetric covariant derivative. It is
important to note, that in the Abelian case the superfield $W^a$
is gauge invariant, so that action (\ref{Regularized_SQED_Action})
will be also gauge invariant.

Quantization of model (\ref{Regularized_SQED_Action}) can be made
by the standard way. For this purpose it is convenient to use the
supergraphs technique, described in book \cite{West} in details,
and to fix the gauge invariance by adding the following terms:

\begin{equation}\label{Gauge_Fixing}
S_{gf} = - \frac{1}{64 e^2}\int d^4x\,d^4\theta\, \Bigg(V D^2 \bar
D^2 \Big(1 + \frac{\partial^{2n}}{\Lambda^{2n}}\Big) V + V \bar
D^2 D^2 \Big(1+ \frac{\partial^{2n}}{\Lambda^{2n}}\Big) V\Bigg),
\end{equation}

\noindent where

\begin{equation}
D^2 \equiv \frac{1}{2} \bar D (1+\gamma_5)D;\qquad \bar D^2 \equiv
\frac{1}{2}\bar D (1-\gamma_5) D.
\end{equation}

\noindent After adding such terms a part of the action, quadratic
in the superfield $V$, will have the simplest form

\begin{equation}
S_{gauge} + S_{gf} = \frac{1}{4 e^2}\int d^4x\,d^4\theta\,
V\partial^2 \Big(1+ \frac{\partial^{2n}}{\Lambda^{2n}}\Big) V.
\end{equation}

\noindent In the Abelian case, considered here, diagrams,
containing ghost loops are absent.

It is well known, that adding of the higher derivative term does
not remove divergences in one-loop diagrams. In order to
regularize them, it is necessary to insert in the generating
functional the Pauli-Villars determinants \cite{Slavnov_Book}.

Due to the supersymmetric gauge invariance the renormalized action
can be presented as

\begin{eqnarray}\label{Renormalized_Action}
&& S_{ren} = \frac{1}{4 e^2} Z_3(e,\Lambda/\mu)\, \mbox{Re}\int
d^4x\,d^2\theta\,W_a C^{ab} \Big(1+
\frac{\partial^{2n}}{\Lambda^{2n}}\Big) W_b
+\nonumber\\
&& \qquad\qquad\qquad\qquad\qquad +
Z(e,\Lambda/\mu)\,\frac{1}{4}\int d^4x\, d^4\theta\, \Big(\phi^*
e^{2V}\phi +\tilde\phi^* e^{-2V}\tilde\phi\Big).\qquad
\end{eqnarray}

\noindent Therefore, the generating functional can be written as

\begin{equation}\label{Modified_Z}
Z = \int DV\,D\phi\,D\tilde \phi\, \prod\limits_i \Big(\det
PV(V,M_i)\Big)^{c_i}
\exp\Big(i(S_{ren}+S_{gf}+S_S+S_{\phi_0})\Big),
\end{equation}

\noindent where the renormalized action $S_{ren}$ is given by Eq.
(\ref{Renormalized_Action}), the gauge fixing action -- by Eq.
(\ref{Gauge_Fixing}) (It is convenient to substitute $e$ by $e_0$
in Eq. (\ref{Gauge_Fixing}), that we will assume below), the
Pauli-Villars determinants are defined by

\begin{equation}\label{PV_Determinants}
\Big(\det PV(V,M)\Big)^{-1} = \int D\Phi\,D\tilde \Phi\,
\exp\Big(i S_{PV}\Big),
\end{equation}

\noindent where

\begin{eqnarray}
&& S_{PV}\equiv Z(e,\Lambda/\mu) \frac{1}{4} \int
d^4x\,d^4\theta\, \Big(\Phi^* e^{2V}\Phi + \tilde\Phi^*
e^{-2V}\tilde\Phi \Big)
+\qquad\nonumber\\
&&\qquad\qquad\qquad\qquad\qquad  + \frac{1}{2}\int
d^4x\,d^2\theta\, M \tilde\Phi \Phi + \frac{1}{2}\int
d^4x\,d^2\bar\theta\, M \tilde\Phi^* \Phi^*,\qquad
\end{eqnarray}

\noindent and the coefficients $c_i$ satisfy conditions

\begin{equation}
\sum\limits_i c_i = 1;\qquad \sum\limits_i c_i M_i^2 = 0.
\end{equation}

\noindent Below we will assume, that $M_i = a_i\Lambda$, where
$a_i$ are some constants. Insertion of Pauli-Villars determinants
allows to cancel remaining divergences in all one-loop diagrams,
including diagrams, containing insertions of counterterms.

The terms with sources are written in the form

\begin{eqnarray}\label{Sources}
&& S_S = \int d^4x\,d^4\theta\,J V + \int d^4x\,d^2\theta\,
\Big(j\,\phi + \tilde j\,\tilde\phi \Big) + \int
d^4x\,d^2\bar\theta\, \Big(j^*\phi^* + \tilde j^*
\tilde\phi^*\Big).
\end{eqnarray}

\noindent Moreover, in generating functional (\ref{Modified_Z}) we
introduced the expression

\begin{equation}
S_{\phi_0} = \frac{1}{4}\int d^4x\,d^4\theta\,\Big(\phi_0^*\,
e^{2V} \phi + \phi^*\, e^{2V} \phi_0 + \tilde\phi_0^*\,
e^{-2V}\tilde\phi + \tilde\phi^*\, e^{-2V}\tilde\phi_0 \Big),
\end{equation}

\noindent where $\phi_0$, $\phi_0^*$, $\tilde\phi_0$ and
$\tilde\phi_0^*$ are scalar superfields. They are some parameters,
which are not chiral or antichiral. In principle, it is not
necessary to introduce the term $S_{\phi_0}$ into the generating
functional, but the presence of the parameters $\phi_0$ is highly
desirable for the investigation of Schwinger-Dyson equations.

In our notations the generating functional for the connected Green
functions is written as

\begin{equation}\label{W}
W = - i\ln Z,
\end{equation}

\noindent and an effective action is obtained by making a Legendre
transformation:

\begin{equation}\label{Gamma}
\Gamma = W - \int d^4x\,d^4\theta\,J V - \int d^4x\,d^2\theta\,
\Big(j\,\phi + \tilde j\,\tilde\phi \Big) - \int
d^4x\,d^2\bar\theta\, \Big(j^*\phi^* + \tilde j^* \tilde\phi^*
\Big),
\end{equation}

\noindent where the sources $J$, $j$ and $\tilde j$ is to be
eliminated in terms of the fields $V$, $\phi$ and $\tilde\phi$,
through solving equations

\begin{equation}
V = \frac{\delta W}{\delta J};\qquad \phi = \frac{\delta W}{\delta
j};\qquad \tilde\phi = \frac{\delta W}{\delta\tilde j}.
\end{equation}

\section{Schwinger-Dyson equation}
\hspace{\parindent} \label{Section_SD}

Following Ref. \cite{SD}, for convenience we first set the
renormalization constant $Z$ equal to 1. The dependence of the
effective action on $Z$ will be restored in the final result.

From generating functional (\ref{Modified_Z}) it is possible to
obtain \cite{SD} the Schwinger-Dyson equations, which can be
graphically written as

\begin{equation}\label{SD_Equation}
\begin{picture}(0,1.8)
\put(-5,0.6){$\Gamma^{(2)}_V =$} \put(0.8,0.6){+} \hspace*{-3.5cm}
\includegraphics[scale=0.88]{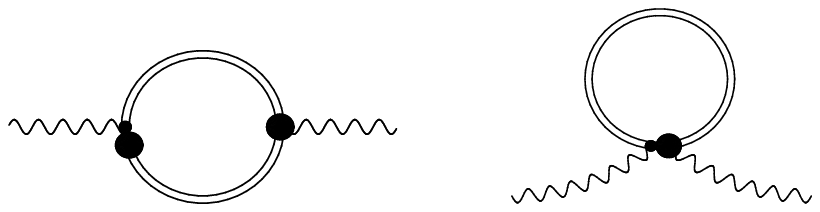}
\end{picture}
\end{equation}

\noindent where $\Gamma^{(2)}_V$ is a two-point Green function of
the gauge field.

The double line denotes the exact propagator ($Z=1$)

\begin{equation}\label{Inverse_Functions}
\Bigg(\frac{\delta^2\Gamma}{\delta\phi_x^*\delta\phi_y}\Bigg)^{-1}
= -  \frac{D_x^2 \bar D_x^2}{4 \partial^2 G} \delta^8_{xy},
\end{equation}

\noindent in which the function $G(q^2)$ is defined by the
two-point Green function as follows:

\begin{equation}\label{Explicit_Green_Functions}
\frac{\delta^2\Gamma}{\delta\phi_x^*\delta\phi_y} = \frac{D_x^2
\bar D_x^2}{16} G(\partial^2) \delta^8_{xy},
\end{equation}

\noindent where $\delta^8_{xy}\equiv
\delta^4(x-y)\delta^4(\theta_x-\theta_y)$, and the lower indexes
denote points, in which considered expressions are taken.

The large circle denotes the effective vertex, which is written as
\cite{SD}

\begin{eqnarray}\label{Vertex2}
&& \frac{\delta^3\Gamma}{\delta
V_x\delta\phi_y\delta\phi^*_{z}}\Bigg|_{p=0} =
\partial^2\Pi_{1/2}{}_x\Big(\bar D_x^2\delta^8_{xy} D_x^2
\delta^8_{xz}\Big) F(q^2) +\nonumber\\
&& \qquad\qquad\quad +\frac{1}{32} q^\mu G'(q^2) \bar
D\gamma^\mu\gamma_5 D_x \Big(\bar D_x^2\delta^8_{xy} D_x^2
\delta^8_{xz}\Big) + \frac{1}{8} \bar D_x^2\delta^8_{xy} D_x^2
\delta^8_{xz}\, G(q^2)
\end{eqnarray}

\noindent due to the Ward identities. Here the strokes denote
derivatives with respect to $q^2$,

\begin{equation}
\Pi_{1/2} = - \frac{1}{16 \partial^2} D^a \bar D^2 D_a = -
\frac{1}{16 \partial^2} \bar D^a D^2 \bar D_a
\end{equation}

\noindent is a supersymmetric transversal projector, and $F(q^2)$
is a function, which can not be defined from the Ward identities.
Here

\begin{eqnarray}
&& D^a \equiv \Big[\frac{1}{2}\bar D (1+\gamma_5)\Big]^a;\qquad
D_a \equiv \Big[\frac{1}{2}(1+\gamma_5) D\Big]_a;\nonumber\\
&& \bar D^a \equiv \Big[\frac{1}{2}\bar D (1 -
\gamma_5)\Big]^a;\qquad \bar D_a \equiv
\Big[\frac{1}{2}(1-\gamma_5) D\Big]_a.
\end{eqnarray}

Two adjacent circles are an effective vertex, consisting of 1PI
diagrams, in which one of the external lines is attached to the
very left edge. Such vertexes are given by \cite{SD}

\begin{equation}\label{Useful_Identities}
\frac{\delta^2\Gamma}{\delta\phi_y \delta \phi_{0z}^*} =
\frac{1}{4} \frac{\delta}{\delta \phi_y}
\exp\Bigg(\frac{2}{i}\frac{\delta}{\delta J_z} + 2 V_z\Bigg)
\phi_z = - \frac{1}{8} G \bar D_y^2\delta^8_{yz}
\end{equation}

\noindent in the case of one external $V$-line (the vertex in the
first diagram of Eq. (\ref{SD_Equation})) and

\begin{eqnarray}\label{Vertex3}
&& \frac{\delta^3\Gamma}{\delta
V_x\delta\phi_y\delta\phi^*_{0z}}\Bigg|_{p=0} = \frac{1}{4}
\frac{\delta}{\delta V_x}\frac{\delta}{\delta \phi_y}
\exp\Bigg(\frac{2}{i}\frac{\delta}{\delta J_z}+2V_z\Bigg)
\phi_z\Bigg|_{p=0} =\nonumber\\
&& = -2 \partial^2\Pi_{1/2}{}_x\Big(\bar D_x^2\delta^8_{xy}
\delta^8_{xz}\Big) F(q^2) + \frac{1}{8} D^a C_{ab} \bar
D_x^2\Big(\bar D_x^2\delta^8_{xy} D_x^b \delta^8_{xz} \Big) f(q^2)
+\vphantom{\frac{1}{2}}\nonumber\\
&&\qquad\qquad\qquad\qquad -\frac{1}{16} q^\mu G'(q^2) \bar
D\gamma^\mu\gamma_5 D_x \Big(\bar D_x^2\delta^8_{xy}
\delta^8_{xz}\Big) -\frac{1}{4} \bar D_x^2\delta^8_{xy}
\delta^8_{xz}\, G(q^2),
\end{eqnarray}

\noindent in the case of two external $V$-lines (in the second
diagram of Eq. (\ref{SD_Equation})). Here $f(q^2)$ is one more
function, which can not be found from the Ward identity. Note,
that such form of the Ward identities solutions can be easily
checked by the method, which is constructed below in this paper.
We will not present here the corresponding calculations, because
they are rather cumbersome.

Our purpose will be calculation of the expression

\begin{equation}
\frac{d}{d\ln\Lambda}\Gamma^{(2)}_V\Bigg|_{p=0},
\end{equation}

\noindent where $p$ denotes the external momentum. \footnote{The
existence of this limit is proved in Ref. \cite{SD}.} In order to
do this, the expressions for the propagators and vertexes given
above are substituted into the Schwinger-Dyson equations. As a
result some expressions for diagrams in Eq. (\ref{SD_Equation})
are obtained. The result \cite{SD} is obtained exactly to all
orders of the perturbation theory. However the second diagram
produces terms, containing the function $f$, which can not be
found from the Ward identities. (In the one-loop approximation
this function is equal 0. It is nontrivial only at two loops.) The
terms, containing the unknown function $F$, are completely
cancelled.

Calculation of contributions, containing the unknown function $f$,
is a main purpose of this paper. After simple transformations the
expression for the second diagram in Eq. (\ref{SD_Equation}) in
the massless case can be written as

\begin{eqnarray}
&& \frac{d}{d\ln\Lambda} \Gamma^{\mbox{diag.\,2}}_V
=\nonumber\\
&& = \frac{d}{d\ln\Lambda} \int d^8x\,d^8y\,d^8z\,V_x V_y
\frac{D_x^2}{4i\partial^2 G} \delta^8_{xz} \frac{\delta}{\delta
V_y} \frac{\delta}{\delta \phi_z}
\exp\Bigg(\frac{2}{i}\frac{\delta}{\delta J_x}+2V_x\Bigg) \phi_x
\Bigg|_{V,\phi=0}.\qquad
\end{eqnarray}

\noindent Then using the identity

\begin{equation}
V_x D_x^2 \delta^8_{xz} = D_x^2 (V_x \delta^8_{xz}) + 2 (D_{a}
V_x) D_x^a \delta^8_{xz} - (D^2 V_x) \delta^8_{xz},
\end{equation}

\noindent and the explicit expression for the vertex function
(\ref{Vertex3}) we find that the sum of two diagrams in Eq.
(\ref{SD_Equation}) (the result of the calculation for the first
diagram is given in Ref. \cite{SD}) is

\begin{eqnarray}\label{Gamma2V}
&& \frac{d}{d\ln\Lambda}\Gamma^{(2)}_V\Bigg|_{p=0} =
\frac{d}{d\ln\Lambda}\int\frac{d^4p}{(2\pi)^4}\frac{d^4q}{(2\pi)^4}
V\partial^2\Pi_{1/2} V\, \frac{1}{2q^2}\frac{d}{dq^2}
\ln(q^2 G^2) +\nonumber\\
&& + \frac{d}{d\ln\Lambda}\int d^8x\,d^8y\,d^8z\,(D_a V_x) V_y
\frac{D_x^a}{2i\partial^2 G} \delta^8_{xz} \frac{\delta}{\delta
V_y} \frac{\delta}{\delta \phi_z}
\exp\Bigg(\frac{2}{i}\frac{\delta}{\delta J_x}+2V_x\Bigg)
\phi_x \Bigg|_{V,\phi=0} -\nonumber\\
&& - \mbox{similar terms with the Pauli-Villars fields}.\qquad
\end{eqnarray}

\noindent The second term will be proportional to

\begin{equation}
\frac{d}{d\ln\Lambda} \int \frac{d^4q}{(2\pi)^4} \frac{f(q^2)}{q^2
G(q^2)}.
\end{equation}

\begin{figure}[h]
\hspace*{6cm}
\begin{picture}(0,0)
\put(0.5,2){$(I_1)^a$} \put(-0.3,-0.4){$D_a V_x$}
\put(2.8,-0.4){$V_y$}
\end{picture}
\includegraphics[scale=0.4]{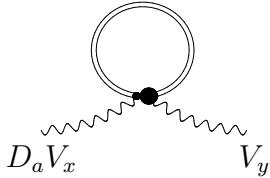}
\newline
\caption{The diagram, giving a contribution to the
$\beta$-function, which can not be found from the Ward
identities.} \label{Figure_Undefined_Diagram}
\end{figure}

\noindent This means, that the second term in equation
(\ref{Gamma2V}) can not be found from the Ward identity. Taking
into account that the variational derivative with respect to a
chiral superfield $\phi_z$ contains $\bar D_z^2$, we easily find
that the contribution of this term can be graphically presented as
a diagram, shown in Fig.\ref{Figure_Undefined_Diagram}. Here the
symbol $(I_1)^a$ (the notations will be explained below) means,
that it is necessary to make a substitution

\begin{equation}
\bar D^2 D^2 \delta^8_{xz} \to 2 \bar D^2 D^a \delta^8_{xz}
\end{equation}

\noindent in the effective propagator. The next section of the
paper is devoted to the partial calculation of this diagram.


\section{Calculation of contributions, which can not be found
from the Ward identities} \label{Section_Green}
\hspace{\parindent}

\begin{figure}[h]
\hspace*{4cm}
\includegraphics[scale=0.4]{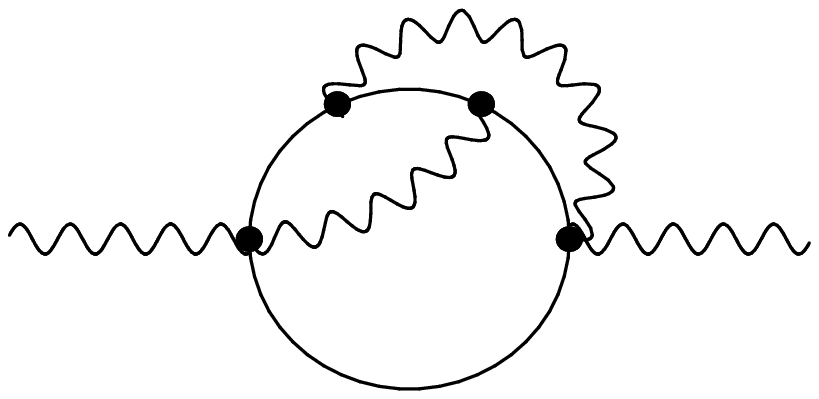}
\hspace*{1cm}
\includegraphics[scale=0.4]{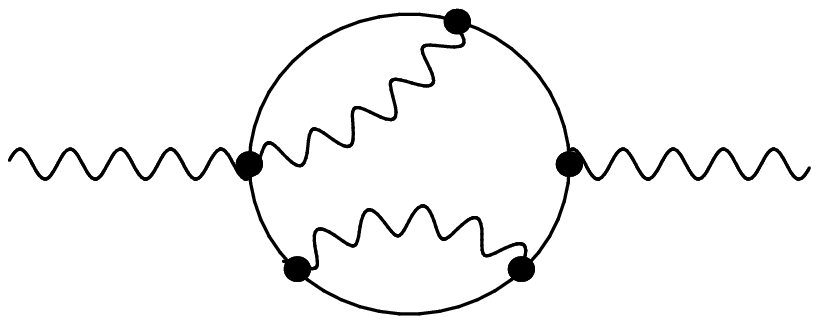}
\newline
\caption{The first of these diagrams belongs to the class of
diagrams, considered in this paper and the second one does not.}
\label{Figure_Examples}
\end{figure}

Let us consider a class of Feynman diagrams, which corresponds to
the effective diagram, presented in Fig.
\ref{Figure_Undefined_Diagram}. It has only one loop of the matter
superfields and the diagram can not be made disconnected cutting
two propagators in this loop. For example, the first diagram in
Fig.\ref{Figure_Examples} belongs to the considered class, while
the second one does not. Then for diagrams of the considered type
we calculate a contribution to the effective diagram, presented in
Fig.\ref{Figure_Undefined_Diagram}.

\begin{figure}[h]
\hspace*{3.5cm}
\begin{picture}(0,0)
\put(0.1,-0.3){1} \put(2.7,-0.3){2} \put(1.4,-0.3){3}
\put(5.15,-0.3){1} \put(7.75,-0.3){2} \put(1.0,1.5){$V$}
\put(7.4,1.5){$V$} \put(4,0.6){+}
\put(0.75,0.05){$\blacktriangleright$}
\put(2.1,0.05){$\blacktriangleright$}
\put(6.5,0.05){$\blacktriangleright$} \put(0.75,0.5){$q$}
\put(2.0,0.5){$q+p$} \put(6.5,0.5){$q$}
\end{picture}
\includegraphics[scale=0.4]{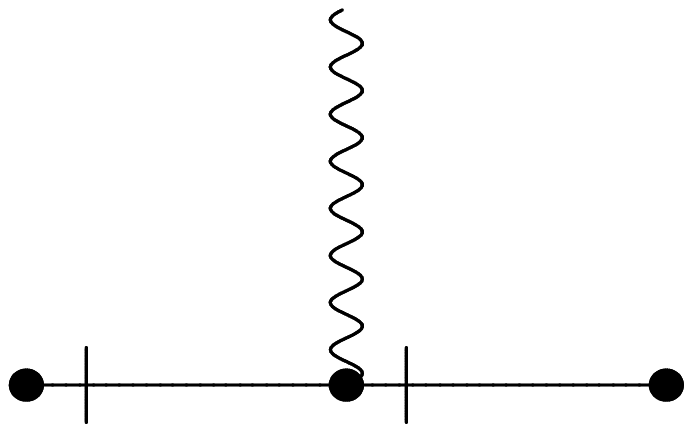}
\hspace*{2cm}
\includegraphics[scale=0.4]{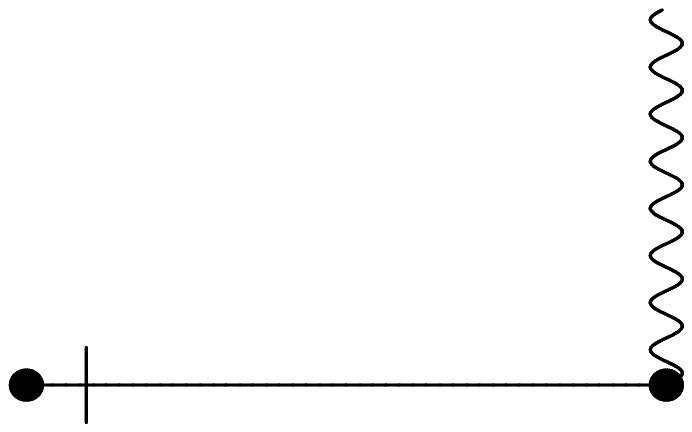}
\newline
\caption{A typical combination of subdiagrams, which appears in
the process of summation of Feynman diagrams.}
\label{Figure_Subdiagrams1}
\end{figure}

For this purpose let us first consider subdiagrams, presented in
Fig. \ref{Figure_Subdiagrams1}. (In our notations if $q_\mu$ is an
incoming momentum, then $\partial_\mu = -iq_\mu$.) Their
contribution is written as

\begin{equation}
\int d^8 x_3 \frac{D_3^2 \bar D_3^2}{4\partial^2}\delta^8_{13}\,
V_3 \frac{\bar D_3^2 D_3^2}{4\partial^2}\delta^8_{23} -
\frac{1}{q^2} \bar D_1^2 D_1^2 \delta^8_{12} V_2
\end{equation}

\noindent In the limit $p\to 0$ in the momentum representation
this expression can be equivalently rewritten as

\begin{eqnarray}\label{Identity1}
&& - \frac{1}{q^2} \bar D_1^2 \delta^8_{12}\, D_2^2 V_2 +
\frac{2}{q^2} \bar D_1^2 D_1^a \delta^8_{12}\, D_{2a} V_2 -
\frac{\bar D_1^2 D_1^a}{2q^4}\delta^8_{12} (\gamma^\mu)_a{}^b \bar
D_{2b} D^2
V_2\,q_\mu +\nonumber\\
&& +  \frac{\bar D_1^2 D_1^2}{q^4} \delta^8_{12}
\Bigg(\frac{1}{2}\partial^2\Pi_{1/2}V_2 -\frac{1}{4} \bar D
\gamma^\mu\gamma_5 D V_2\, q_\mu \Bigg),
\end{eqnarray}

\noindent where the anticommuting relations for the covariant
derivatives were used. This can be graphically interpreted as
follows: Diagrams in Fig. \ref{Figure_Subdiagrams1} in the limit
$p\to 0$ are equivalent to the sum of diagrams, in which the
external line is attached to point 2 and corresponds to the
expressions $D^2 V,\ldots, \bar D\gamma^\mu\gamma_5 D V$, while
the line, connecting points $1$ and $2$, gives the contribution $-
\bar D^2_1\delta^8_{12}/q^2,\ldots,-q_\mu/q^4 \bar D_1^2
D_1^2\delta^8_{12}$ respectively.

Because the considered diagram, presented in Fig.
\ref{Figure_Undefined_Diagram}, already contains an external line
with $D_a V$, the terms, which contain only the first degrees of
$D^a$ will be essential for us. (The other terms are 0 in the
limit $p^\mu\to 0$.) These terms are evidently the ones containing
$D_a V_2$ and $\bar D \gamma^\mu\gamma_5 D V_2$.

Moreover, we need some identities, which allow to carry the
external line from one vertex to another through a propagator. In
order to make the formulas not too large, it is convenient to use
the following notations:

\begin{eqnarray}
&& (I_0) \equiv \frac{1}{4} \bar D_1^2 D_1^2\delta^8_{12};\qquad
(I_1)^a \equiv \frac{1}{2} \bar D_1^2 D_{1}^a \delta^8_{12};
\qquad (\bar I_1)^a \equiv \frac{1}{2} \bar D_1^a D_1^2 \delta^8_{12};
\nonumber\\
&& (I_2) \equiv \frac{1}{4} \bar D_1^2 \delta^8_{12};\qquad\quad\,
(\bar I_2) \equiv \frac{1}{4} D_1^2 \delta^8_{12};\qquad\quad\ \
\, (I_2)^{ab}
\equiv \bar D_1^a D_1^b\delta^8_{12};\nonumber\\
&& (I_3)^a \equiv \frac{1}{2} D_1^a \delta^8_{12}; \qquad\ \ \,
(\bar I_3)^a \equiv \frac{1}{2} \bar
D_1^a\delta^8_{12};\qquad\quad\ (I_4) \equiv
\frac{1}{4}\delta^8_{12}.
\end{eqnarray}

\noindent Then, for example, if $X$ is an even element of the
Grassmann algebra

\begin{eqnarray}\label{Transfer_Identities}
&&\hspace*{-3mm} \frac{1}{4}\bar D_1^2 D_1^2 \delta^8_{12}\, X_2 =
\frac{1}{4}\bar D_1^2 D_1^2\Big(X_1 \delta^8_{12}\Big) = X_1 (I_0)
+ D^a X_1 (I_1)_a + \bar D^a X_1 (\bar I_1)_a + D^2 X_1 (I_2) +\nonumber\\
&&\hspace*{-3mm} + \bar D^2 X_1 (\bar I_2) - \bar D^a D^b X_1
(I_2)_{ab} + \bar D^2 D^a X_1 (I_3)_a + \bar D^a D^2 X_1 (\bar
I_3)_a + \bar D^2 D^2 X_1 (I_4).
\end{eqnarray}

\noindent (Let us remind, that $X_1$ denotes, that the field $X$
is taken in point 1, and $X_2$ -- in point 2.) Other similar
identities can be easily found.

The idea of Feynman diagrams summation is that we first use
identity (\ref{Identity1}) and in all diagrams remove vertexes,
with one attached external line of the field $V$, but with no
attached internal $V$-lines. After this we carry one of the
external lines of the superfield $V$ around the loop of the matter
superfields until the point, to which the other external line is
attached, by using identity (\ref{Transfer_Identities}) and other
similar identities. As a simple exercise to application of this
method it is possible to verify Eqs. (\ref{Vertex2}) and
(\ref{Vertex3}), which are the solutions of the Ward identities
(in this case there are no limitations to the form of diagrams),
and also to obtain explicit expressions for the unknown functions
$f$ and $F$ as some sums of diagrams. Due to the lack of place we
will not present the detailed solution of this problem. We will
apply the described method directly to the calculation of the
diagram, presented in Fig. \ref{Figure_Undefined_Diagram}.

\begin{figure}[h]
\hspace*{2.9cm}
\begin{picture}(0,0)
\put(-0.2,0.1){$D_a V$} \put(3.2,0.1){$D^b V$}
\put(2.4,-0.3){$(I_1)_b$} \put(0.7,1.3){$(I_1)^a$}
\put(5.2,0.1){$D_a V$} \put(8.6,0.9){$\bar D \gamma^\mu\gamma_5 D
V$} \put(7.8,-0.3){$q_\mu/4q^2 (I_0)$} \put(6.1,1.3){$(I_1)^a$}
\put(4.4,0.6){$+$}
\end{picture}
\includegraphics[scale=0.4]{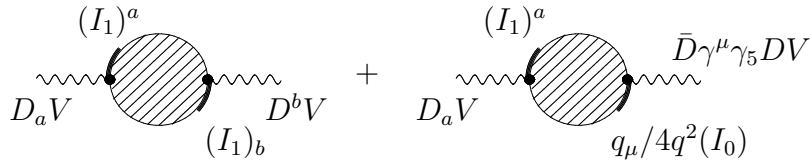}
\hspace*{2cm}
\includegraphics[scale=0.4]{ins3.eps}
\newline
\caption{The result of application identity (\ref{Identity1}) to a
diagram, presented in Fig. \ref{Figure_Undefined_Diagram}.}
\label{Figure_X}
\end{figure}

First we use identity (\ref{Identity1}) and remove vertexes, with
one attached external line of the field $V$, but with no attached
internal $V$-lines. The result can be graphically presented as a
sum of two diagrams, shown in Fig. \ref{Figure_X}. (All other
diagrams give zero result in the limit $p\to 0$.) In Fig.
\ref{Figure_X} and in the subsequent figures the hatched circle
denotes a set of all internal lines of a diagram. {\it External
lines are attached only to points, with at least one more attached
$V$-line.} Bold lines in the figure denote propagators, in which
$(I_0)$ is changed by $(I_1)^a$ e t.c. In two considered diagrams
such propagators are attached directly to the external lines.

To perform the further calculations it is convenient to use the
following notations for the Feynman diagrams:

Let us consider a loop of the matter superfield and a fixed point
on it. Let us go around this loop in some direction (for example,
clockwise) and look, what expressions are encountered in places,
where the propagators of the matter superfields were originally.

If a sequence of ordinary propagators

\begin{equation}
\frac{\bar D_1^2 D_1^2}{4 q^2}\delta^8_{12}
\end{equation}

\noindent is encountered, we will denote it by the symbol $*$.
However other expressions can be encountered in some places, e.f.,
$(I_1)_a/q^2$ -- $(I_4)/q^2$. In this case we write the
corresponding $(I)$ in the expression for the diagram, omitting a
factor $1/q^2$ for brevity. The momentum of the considered
propagator we will always denote by $q$ (it is not a loop
momentum). If an expression like

\begin{equation}
q^\mu \frac{\bar D_1^2 D_1^2}{4 q^4}\delta^8_{12},
\end{equation}

\noindent is encountered we again multiply it by $q^2$ and write
the result as $q^\mu/q^2 (I_0)$. At last, if an external line is
encountered, we will write an expression, corresponding to this
external line. In this way we pass around the loop until the
starting point.

\begin{figure}[h]
\hspace*{6.3cm}
\begin{picture}(0,0)
\put(-0.5,0.1){$\bar D^2 V$} \put(3.2,0.1){$D^2 V$}
\put(2.3,-0.3){$(\bar I_1)^a$} \put(0.6,1.3){$(I_1)_b$}
\put(4.1,0.5){$(\gamma^\mu)_a{}^b$}
\put(1.65,-0.08){$\blacktriangleright$}
\put(1.65,1.12){$\blacktriangleleft$} \put(1.7,-0.4){$q$}
\put(2.0,1.4){$q+p$}
\end{picture}
\includegraphics[scale=0.4]{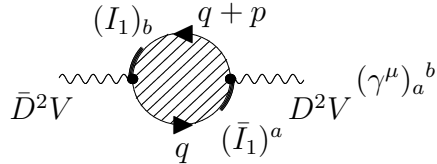}
\newline
\caption{The effective diagram for obtaining identity
(\ref{Equality2}).} \label{Figure_Diagram_Identity12}
\end{figure}

Thus to any Feynman diagram we set to the correspondence a special
symbol. We will assume summation over all diagrams with the same
symbol. For example, the expression $* D^2 V (\bar I_1)^a
* \bar D^2 V (I_1)_b (\gamma^\mu)_a{}^b$ denotes the sum of all
diagrams, which have a structure, corresponding to the diagram in
Fig. \ref{Figure_Diagram_Identity12}. It is worth noting, that
there are no limitations to possible positions of the insertions
$(I)$ for the diagrams of the considered class, while for all
other diagrams positions of some insertions are not in general
completely arbitrary.

For calculation of diagrams in Fig.\ref{Figure_X} by identities,
similar to (\ref{Transfer_Identities}), in both diagrams we carry
the right external line clockwise until a point to which the left
external line is attached. It is easy to see, that only the terms

\begin{eqnarray}\label{Considered_Terms}
&&  - 4 \frac{d}{d\ln\Lambda} \Bigg( (\gamma^\mu)_a{}^b \Big( *
\frac{q^\mu}{2q^2} (\bar I_1)^a * (I_1)_b + * \frac{q^\mu}{2q^2}
(I_0) * (\bar I_1)^a * (I_1)_b \Big) +\nonumber\\
&& + 2 * (I_3)^a * (I_1)_a  + 2 * (I_1)^a * (\bar I_2) * (I_1)_a -
* (I_2)^{ab} * (\bar I_1)_a * (I_1)_b -\vphantom{\frac{1}{2}}\nonumber\\
&& - * (I_1)^a * (\bar I_1)^b * (\bar I_1)_{b} * (I_1)_a\Bigg)
V\partial^2 \Pi_{1/2} V.
\end{eqnarray}

\noindent are not finally equal to 0 in the limit $p^\mu\to 0$.
(In this case such sequence of writing is chosen: from the point
to which the right external line is attached, clockwise.)

\begin{figure}[h]
\hspace*{0.8cm}
\begin{picture}(0,0)
\put(1.7,-0.1){$\bar D^2 V$} \put(5.6,-0.1){$D^2 V$}
\put(4.7,0.5){$q^\mu/q^2 (I_0)$} \put(7.1,-0.1){$\bar D^2 V$}
\put(11.0,-0.1){$D^2 V$} \put(10.1,0.5){$(I_2)_b{}^a$}
\put(11.4,0.8){$(\gamma^\mu)_a{}^b$}
\end{picture}
\hspace*{2.4cm}
\includegraphics[scale=0.4]{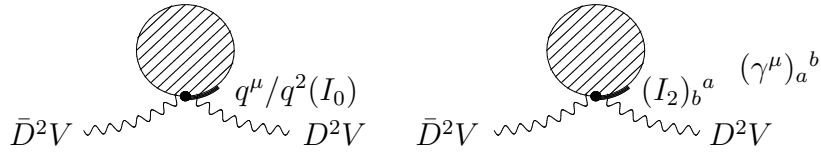}
\hspace*{2.4cm}
\includegraphics[scale=0.4]{ins2.eps}
\newline
\caption{Effective diagrams for obtaining identity
(\ref{Equality3}).} \label{Figure_Diagram_Identity3}
\end{figure}

\begin{figure}[h]
\hspace*{3.35cm}
\begin{picture}(0,0)
\put(1.7,-0.1){$\bar D^a V$} \put(5.6,-0.1){$D^2 V$}
\put(4.7,0.5){$(\bar I_1)_a$} \put(4.7,1.7){$(I_1)_b$}
\put(2.4,1.2){$(I_1)^b$}
\end{picture}
\hspace*{2.4cm}
\includegraphics[scale=0.4]{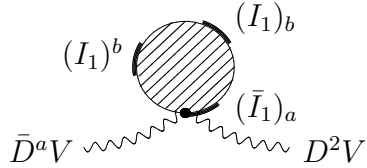}
\newline
\caption{Effective diagram for obtaining identity
(\ref{Identity5}).} \label{Figure_Diagram_Identity5}
\end{figure}

In order to prove, that expression (\ref{Considered_Terms}) is 0,
we need some identities, which can be found from the consideration
of diagrams, presented in Fig. \ref{Figure_Diagram_Identity12} --
\ref{Figure_Diagram_Identity5}.

As an example we investigate a diagram, presented in Fig.
\ref{Figure_Diagram_Identity12}. The expression for it we will
denote $F^\mu(p)$ for the definiteness. We will calculate this
expression by two ways: carrying in the considered diagram a right
external line around the loop of the matter superfield clockwise
and counter-clockwise until a point, to which another external
line is attached. In the obtained expressions we will keep only
terms of the first order in $p$. Such terms are produced, first,
from expansions of the propagators momentums, and, second, from
action of the operators $D$ and $\bar D$ to external lines. The
result of the described procedure is two expressions for

\begin{equation}
\frac{\partial}{\partial p^\nu} F^\mu(p)\Bigg|_{p=0} p^\nu,
\end{equation}

\noindent from which it is easy to construct two different
expressions for $\partial F^\mu/\partial p^\mu|_{p=0}$. Equating
them we obtain the identity

\begin{eqnarray}\label{Equality2}
&& (\gamma^\mu)_a{}^b \Big( * (\bar I_1)^a
* \frac{q^\mu}{q^2} (I_1)_b + * (\bar I_1)^a * (I_1)_b
* \frac{q^\mu}{q^2} (I_0) + * \frac{q^\mu}{q^2} (\bar I_1)^a
* (I_1)_b
+\nonumber\\
&& + * (\bar I_1)^a * \frac{q^\mu}{q^2} (I_0) * (I_1)_b \Big) + 8
* (\bar I_3)^a * (\bar I_1)_a - 4 * (I_2)^{ab} * (I_1)_b *
(\bar I_1)_a   -\qquad\vphantom{\frac{1}{2}}\nonumber\\
&& - 2 * (I_2)^{ab} * (\bar I_1)_a * (I_1)_b - 4 * (\bar I_1)^a *
(I_1)^{b} * (\bar I_1)_a * (I_1)_b =
0.\qquad\vphantom{\frac{1}{2}}
\end{eqnarray}

Similar consideration of diagrams, presented in Fig.
\ref{Figure_Diagram_Identity3} \footnote{In the first diagram in
Fig. \ref{Figure_Diagram_Identity3} we calculate terms of the
first order in the momentum $p$ by two ways: carrying the line
with $D^2 V$ and the line with $\bar D^2 V$ around the loop. In
the second diagram the line with $\bar D^2 V$ is carried around
the loop until the starting point and terms of the first order in
the external momentum $p$ are equated to zero.}, gives the
identity

\begin{eqnarray}\label{Equality3}
&& (\gamma^\mu)_a{}^b \Big( * (\bar I_1)^a
* \frac{q^\mu}{q^2} (I_1)_b + * (\bar I_1)^a * (I_1)_b
* \frac{q^\mu}{q^2} (I_0) + * \frac{q^\mu}{q^2} (\bar I_1)^a
* (I_1)_b
+\nonumber\\
&& + * (\bar I_1)^a * \frac{q^\mu}{q^2} (I_0) * (I_1)_b \Big) + 8
* (\bar I_3)^a * (\bar I_1)_a - 2 * (I_2)^{ab} * (\bar I_1)_a *
(I_1)_b  = 0,\qquad\vphantom{\frac{1}{2}}
\end{eqnarray}

\noindent where we took into account, that due to symmetry
properties of the sum of Feynman diagrams

\begin{eqnarray}
&& * (I_3)^a * (I_1)_a = * (\bar I_3)^a * (\bar I_1)_a;\nonumber\\
&& * (I_1)^a  * (I_1)_a * (\bar I_2) = * (\bar I_1)^a  * (\bar
I_1)_a * (I_2).\vphantom{\frac{1}{2}}
\end{eqnarray}

\noindent Comparing Eqs. (\ref{Equality2}) and (\ref{Equality3}),
we find

\begin{equation}\label{Identity}
* (I_2)^{ab}  * (I_1)_b * (\bar I_1)_a +
* (\bar I_1)^b * (I_1)^a * (\bar I_1)_{b} * (I_1)_a = 0.
\end{equation}

\noindent Finally, one more identity is obtained from the diagram,
presented in Fig. \ref{Figure_Diagram_Identity5} in the limit
$p\to 0$ after carrying the line with $D^2 V$ around the loop. The
original diagram is evidently 0, but the carrying of the line
gives terms, proportional to $V\partial^2 \Pi_{1/2} V$. Their sum
should be 0. Calculating these terms, we come to the equality

\begin{eqnarray}\label{Identity5}
&& 0 = 2 * (\bar I_1)^a * (\bar I_1)_a * (I_1)^b * (I_1)_b - *
(I_1)^a * (\bar I_1)^b * (I_1)_a * (\bar I_1)_b +\nonumber\\
&& + *(I_2)^{ab} * (\bar I_1)_a * (I_1)_b  - *(I_2)^{ab} * (I_1)_b
* (\bar I_1)_a - 4 * (\bar I_2) * (I_1)^b * (I_1)_b.
\vphantom{\frac{1}{2}}\qquad
\end{eqnarray}

\noindent From identities (\ref{Equality3}), (\ref{Identity}) and
(\ref{Identity5}) we find

\begin{eqnarray}
&& 0 = (\gamma^\mu)_a{}^b \Big( * (\bar I_1)_b
* \frac{q^\mu}{4q^2} (I_1)^a + * (\bar I_1)_b * (I_1)^a
* \frac{q^\mu}{4q^2} (I_0) + * \frac{q^\mu}{4q^2} (\bar I_1)_b
* (I_1)^a
+\nonumber\\
&& + * (I_1)_b * \frac{4q^\mu}{q^2} (I_0) * (I_1)^a \Big) + 2 *
(\bar I_2) * (\bar I_1)^a  * (\bar I_1)_a
+ 2 * (\bar I_3)^a * (\bar I_1)_a -\nonumber\\
&& - * (I_2)^{ab} * (\bar I_1)_a * (I_1)_b
 - * (\bar I_1)^b * (\bar I_1)_{b}
* (I_1)^a * (I_1)_a.\vphantom{\frac{1}{2}}
\end{eqnarray}

\noindent Comparing this equality with expression
(\ref{Considered_Terms}), we find, that the sum of the considered
diagrams will be proportional to

\begin{equation}\label{Last_Expression}
(\gamma^\mu)_a{}^b\Big(* (\bar I_1)_b * (I_1)^a *
\frac{q^\mu}{q^2} (I_0) +
* (I_1)^a * (\bar I_1)_b * \frac{q^\mu}{q^2} (I_0)\Big).
\end{equation}

\noindent (All other terms are cancelled due to the symmetries of
the sum of Feynman diagrams.)

It is easy to see, that for the diagrams of the considered class
this expression is 0. Really, due to the identity

\begin{equation}
\delta^8_{12} X_2 = X_1 \delta^8_{12}
\end{equation}

\begin{figure}[h]
\hspace*{3cm}
\begin{picture}(0,0)
\put(3.7,-0.1){1} \put(5,-0.1){4} \put(3.7,1.2){2} \put(5,1.2){3}
\end{picture}
\hspace*{3.5cm}
\includegraphics[scale=0.4]{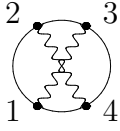}
\newline
\caption{The diagram, illustrating that expression
(\ref{Last_Expression}) is 0.} \label{Figure_Passes}
\end{figure}

\noindent an external line can be carried through a propagator of
the field $V$. It is easy to see, that in any diagram of the
considered class there is a pass (which includes carrying over
photon lines), which allows to carry an external line so that the
passing sequence of internal lines with $(I_1)^a$ and $(\bar
I_1)_b$ would be opposite to the original one. Then the diagram
with replaced $(\bar I_1)_b$ and $(I_1)^a$ is effectively
obtained. For example, in Fig. \ref{Figure_Passes} the original
diagram corresponds to the sequence $1\to 2\to 3\to 4\to 1$, while
the pass $1\to 2\to 4\to 1\to 3\to 4\to 2\to 3\to 1$ is also
possible. Therefore, expression (\ref{Last_Expression}) in the
considered class of diagrams is 0 after summing all diagrams.

Thus the result of the above calculations for diagrams of the
considered class can be presented as the identity

\begin{equation}\label{New_Identity}
\frac{d}{d\ln\Lambda} \int \frac{d^4q}{(2\pi)^2}\frac{f(q^2)}{q^2
G(q^2)} = 0,
\end{equation}

\noindent where the function $f$ was defined by Eq.
(\ref{Vertex3}), and the function $G$ -- by Eq.
(\ref{Explicit_Green_Functions}). Note, that the explicit three-
\cite{ThreeLoop} and four-loop \cite{Pimenov} calculations show,
that identity (\ref{New_Identity}) is valid for arbitrary
diagrams, but their consideration appears to be a bit more
complicated. Note, that this identity is not a consequence of the
Ward identities. It is found by explicit summation of Feynman
diagrams.

Let us write the quantum corrections to the effective action,
corresponding to the two-point Green function of the gauge field,
in the form

\begin{equation}
\Gamma^{(2)}_V = - \frac{1}{16\pi} \int
\frac{d^4p}{(2\pi)^4}\,V(-p)\,\partial^2\Pi_{1/2} V(p)\,
d_0^{-1}(\alpha_0,\Lambda/p),
\end{equation}

\noindent where $d_0$ is some function (the index $0$ of the
function $d$ means, that $Z=1$ so far). Then taking into account
results of Ref. \cite{SD} (see also Eq. (\ref{Gamma2V})) from the
obtained identity we find

\begin{equation}\label{Proposal}
\frac{d}{d\ln\Lambda} d_0^{-1}(\alpha,\mu/p) \Bigg|_{p=0} = -16\pi
\frac{d}{d\ln\Lambda}\int\frac{d^4q}{(2\pi)^4}
\Bigg(\frac{1}{2q^2}\frac{d}{dq^2} \ln(q^2 G^2) - (PV)\Bigg).
\end{equation}

\noindent where $(PV)$ denotes a contribution of the Pauli-Villars
fields. Thus, the proposal, made in \cite{SD}, has been proved in
the massless case for diagrams of the considered type. For the
Pauli-Villars fields the similar proposal is

\begin{eqnarray}\label{PV_Proposal}
(PV) = \sum\limits_i c_i \frac{1}{2q^2}\frac{d}{dq^2}
\Bigg(\ln\Big(q^2 G_{PV}^2 + M_i^2 J_{PV}^2\Big)+ \frac{M_i^2
J_{PV}}{q^2 G_{PV}^2 + M_i^2 J_{PV}^2}\Bigg),
\end{eqnarray}

\noindent where the functions $G_{PV}$ and $J_{PV}$ are defined by
the two-point Green functions for the Pauli-Villars fields. The
proof of this proposal requires similar summation of diagrams in
the massive case, which is essentially more complicated
technically. However, the results of explicit three-loop
calculations confirm this statement and we hope to give its
general proof in future papers.


\section{Two-point Green function of the gauge field.}
\label{Section_Anomaly_Puzzle} \hspace{\parindent}

Let us analyze consequences of Eqs. (\ref{Proposal}) and
(\ref{PV_Proposal}). Due to the identity, partially proven in this
paper, the integrals, defining the two-point Green function of the
gauge field, are reduced to integrals of total derivatives in the
four-dimensional spherical coordinates:

\begin{eqnarray}\label{Limit}
&& \frac{d}{d\ln\Lambda} d^{-1}(\alpha_0,\Lambda/p) \Bigg|_{p=0} =
- \frac{d}{d\ln\Lambda} \frac{1}{2\pi}\Bigg\{\ln G^2 -
\sum\limits_i
c_i \ln\Big(M_i^2 J_{PV}^2\Big)\Bigg\}\Bigg|_{q=0} =\nonumber\\
&& = \frac{1}{\pi}\Bigg(1-\frac{d\ln G}{d\ln
\Lambda}\Bigg)\Bigg|_{q=0} = \frac{1}{\pi}
\frac{d}{d\ln\Lambda}\Bigg(\ln\frac{\Lambda}{p} - \ln
G(\alpha_0,\Lambda/p) \Bigg)\Bigg|_{p=0}.
\end{eqnarray}

\noindent Performing the differentiation it is also necessary to
take into account the dependence of $\alpha_0$ on $\Lambda$. More
detailed explanation of this result is given in Ref. \cite{SD}.
Existence of the considered limit for finite $\Lambda$ is also
proved in Ref. \cite{SD}. Therefore, as a consequence of the Eq.
(\ref{Limit}) we find

\begin{equation}
d_0^{-1}(\alpha_0,\Lambda/p) = \frac{\pi}{\alpha_0} +
\frac{1}{\pi} \ln\frac{\Lambda}{p} - \ln G(\alpha_0,\Lambda/p) +
\mbox{const}.
\end{equation}

Let us now restore the dependence of the effective action on the
renormalization constant $Z$. It is easy to see from the Feynman
rules, that for this purpose it is sufficient to make a
substitution $M_i\to M_i/Z$ in contributions of Pauli-Villars
fields. Then from Eq. (\ref{PV_Proposal}) we easily find

\begin{equation}
d^{-1}(\alpha,\mu/p) = \frac{\pi}{\alpha_0} + \frac{1}{\pi}
\ln\frac{\Lambda}{p} - \ln (ZG) + \mbox{const},
\end{equation}

\noindent where the function $d$ is defined exactly as $d_0$, but
the dependence of the effective action on $Z$ should be taken into
account. The expression $ZG$ is finite according to the definition
of the renormalization constant $Z$. Therefore, in order to make
this expression finite, it is necessary to cancel only the
one-loop divergence. For this purpose the renormalization constant
$Z_3$ should be chosen so that

\begin{equation}\label{Bare_Coupling_Constant}
\frac{1}{e^2} Z_3\Big(e,\Lambda/\mu\Big) = \frac{1}{e^2} -
\frac{1}{4\pi^2}\ln\frac{\Lambda}{\mu},
\end{equation}

\noindent where $\mu$ is a normalization point. Therefore the
final expression for the effective action (without the gauge
fixing terms) in the massless case can be written as

\begin{equation}\label{Effective_Action}
\Gamma_V^{(2)} = \int \frac{d^4p}{(2\pi)^4}\,W_a(-p)\,C^{ab}
W_b(p) \,\Bigg[\frac{1}{4e^2} + \frac{1}{16\pi^2} \ln\frac{\mu}{p}
- \frac{1}{16\pi^2}\ln (Z G) + \mbox{const}\Bigg].
\end{equation}

According to this formula the Gell-Mann-Low function, defined by
Eq. (\ref{GL_Function}), is written as

\begin{equation}
\beta(\alpha) =
\frac{\alpha^2}{\pi}\Big(1-\gamma(\alpha)\Big),\quad\mbox{where}\quad
\gamma \equiv \frac{\partial}{\partial\ln x}\,\ln(ZG)\Bigg|_{x=1},
\end{equation}

\noindent has corrections in all orders of the perturbation theory
and coincides with the exact NSVZ $\beta$-function.

Note, that with the higher derivative regularization the scheme,
in which the exact NSVZ $\beta$-function is obtained, is defined
as follows: The renormalization of the operator $W_a C^{ab} W_b$
is made without adding finite counterterms in two- and more loops.
(There are no divergencies in the corresponding Green function in
that loops.) Finite counterterms, which can be added for the
renormalization of the two-point Green function of the matter
superfield, can be arbitrary.


\section{Conclusion}
\label{Section_Conclusion}
\hspace{\parindent}

Summation of Feynman diagrams, defining the two-point Green
function of the gauge field in the limit $p\to 0$ in the massless
$N=1$ supersymmetric electrodynamics, is partially made by
Schwinger-Dyson equations and Ward identities. Diagrams, which can
not be summed by this way, are non-planar and give a nontrivial
contribution starting from the three-loop approximation.
Calculation of their sum exactly to all orders of the perturbation
theory appears to be rather complicated technically and was
performed in this paper for a certain (sufficiently large) class
of diagrams. This class of diagrams is chosen because for it the
technique of calculations is simpler. The other diagrams probably
can be investigated similarly. It is most important, that there is
a summation algorithm, constructed in this paper.

Note, that the result of diagrams summation, which is expressed by
Eq. (\ref{New_Identity}) and can be easily guessed from explicit
calculations, is a new identity, which is not reduced to the Ward
identities. The results of this paper and Ref. \cite{SD} show,
that this identity is closely connected with relation
(\ref{Effective_Action}) for the two-point Green functions, from
which the exact NSVZ $\beta$-function is obtained. Then there are
some questions: What is a true reason of this identity? Is it a
consequence of some symmetry of the theory? Can the obtained
identity together with relation for the Green functions
(\ref{Effective_Action}) be included in a set of identities? So
far we have no answers to these questions.

Moreover, there are some purely technical problems: It is a proof
of identity (\ref{New_Identity}) for diagrams, which were not
considered in this paper, and of the similar identity in the
massive case (and in particular for diagrams with the
Pauli-Villars fields). The existence of such identity can be
rather confidently suggested from the explicit calculations.
Certainly, it would be also desirable to formulate all the results
in terms of the generating functional, but so far we did not
manage to do this.

The results, obtained by summation of the perturbation theory
series, means, that Gell-Mann-Low function (\ref{GL_Function})
coincides with the exact NSVZ $\beta$-function. (The
renormalization of the operator $W_a C^{ab} W_b$ is exhausted at
the one-loop.)

Also it would be interesting to perform the similar investigation
for the supersymmetric Yang-Mills theory, using higher covariant
derivative regularization or simpler similar noninvariant
regularization. In the latter case it is necessary to use a
special subtraction scheme, which allows to restore the gauge
invariance \cite{Slavnov1,Slavnov2}. Such scheme was proposed in
Ref. \cite{SlavnovStepan1} for Abelian supersymmetric theories and
in Ref. \cite{SlavnovStepan2} for the non-Abelian theories.

\bigskip
\bigskip

\noindent {\Large\bf Acknowledgements.}

\bigskip

\noindent This work was supported by RFBR grant No 02-01-00126.



\begin{thebibliography}{100}

\bibitem{NSVZ_Instanton}
{\it V.Novikov, M.Shifman, A.Vainstein, V.Zakharov}, Phys.Lett.
{\bf 166B}, (1985), 329.

\bibitem{Siegel}
{\it W.Siegel}, Phys.Lett. {\bf 84 B}, (1979), 193.

\bibitem{Tarasov}
{\it O.V.Tarasov, V.A.Vladimirov}, Phys.Lett. {\bf 96 B}, 94,
(1980).

\bibitem{North}
{\it I.Jack, D.R.T.Jones, C.G.North}, Nucl.Phys. {\bf B 486}, 479,
(1997).

\bibitem{hep}
{\it A.Soloshenko, K.Stepanyantz}, Two-loop renormalization of
$N=1$ supersymmetric electrodynamics, regularized by higher
derivatives, hep-th/0203118.

\bibitem{tmf2}
{\it A.Soloshenko, K.Stepanyantz}, Theor.Math.Phys., {\bf 134},
(2003), 377.

\bibitem{ThreeLoop}
{\it A.Soloshenko, K.Stepanyantz}, Theor.Math.Phys. {\bf 140},
(2004), 1264.

\bibitem{Pimenov}
{\it A.Pimenov, K.Stepanyantz}, in preparation.

\bibitem{Slavnov}
{\it A.A.Slavnov}, Theor.Math.Phys. {\bf 23}, (1975), 3.

\bibitem{Bakeyev}
{\it T.Bakeyev, A.Slavnov}, Mod.Phys.Lett. {\bf A11}, (1996),
1539.

\bibitem{SD}
{\it K.V.Stepanyantz}, Theor.Math.Phys. {\bf 142}, (2005), 29.

\bibitem{West}
{\it P.West}, Introduction to supersymmetry and supergravity,
World Scientific, 1986.

\bibitem{Slavnov_Book}
{\it L.D.Faddeev, A.A.Slavnov}, Gauge fields, introduction to
quantum theory, second edition, Benjamin, Reading, 1990.

\bibitem{Slavnov1}
{\it A.A.Slavnov}, Phys.Lett. {\bf B 518}, (2001), 195.

\bibitem{Slavnov2}
{\it A.A.Slavnov}, Theor.Math.Phys. {\bf 130}, (2002), 1.

\bibitem{SlavnovStepan1}
{\it A.A.Slavnov, K.V.Stepanyantz}, Theor.Math.Phys., {\bf 135},
(2003), 673.

\bibitem{SlavnovStepan2}
{\it A.A.Slavnov, K.V.Stepanyantz}, Theor.Math.Phys., {\bf 139},
(2004), 599.

\end{thebibliography}
\end{document}